\documentclass[aps,showpacs,prl,toolkits,nofootinbib,preprint]{revtex4}

\usepackage{mathrsfs}

\usepackage{amsmath}

\usepackage{graphicx}

\usepackage{color}

\usepackage{epsfig}

\usepackage{subfigure}

\begin{document}

\title{Is the Froissart bound relevant for the total $pp$ cross section\\ at $s$=(14 TeV)$^2$?}

\author{Shmuel Nussinov}

\email{nussinov@post.tau.ac.il} \affiliation{Tel Aviv University, Sackler School Faculty of Sciences, \\Ramat Aviv, Tel Aviv 69978, Israel}

\begin{abstract}

  The Froissart bound limmits the asymptotic  $s\rightarrow\infty$ behavior of crossections by
 $(\pi/t_0)\;ln ^2 (s/{(s_0)} $ where  $ t_0$  is the lightest exchanged particle , or more generally the nearest ssingularity,
 in the t channel. We suggest that in comparing this bound with data at energies less than those of LHC, gluebaall masses raather than the small pion mass should be used for $(t_0)^{1/2}$.

\end{abstract}


\maketitle

    The upper bound on total cross sections at asymptotic
energies~\cite{Fr,Ma,Lu,He}:

 $\sigma_{tot} (s) < (\pi/{t_0}) ln ^2 (s/{s_0})$
is well known. Its proof uses:

  1) The unitarity bound:

\begin{equation}
                     0 < a_l < 1
\label{unitarityboundEq2}
\end{equation}

on the imaginary parts of partial waves in the expansion of the elastic two particle scattering amplitude:

\begin{equation}
     {\rm Im}([f(k,\theta)]) = \Sigma (2l+1) \, a_l \, P_l(cos\theta)
\label{elasticabEq3}
\end{equation}

with $\theta$ the scattering angle in the "center of mass" Lorentz frame, $k$ the three momentum of each particle, and $W=s^{1/2}$
the total energy in the CM  with $s=W^2=4(k^2+m^2)$ and $|f|^2$
is the differential elastic cross section for momentum transfer $t= -2k^2(1-{\rm cos}(\theta))$.

 2) Analyticity of the $2 \rightarrow 2$ scattering amplitude in t for $t<t_0$ with
 the $(t_0)^{1/2}$ the lightest mass exchangeable in the t channel. This implies
 that the partial wave expansion (\ref{elasticabEq3}) converges inside the "Martin-Lehman"
 ellipse~\cite{Leh} with foci +1 and -1 and with semi-major axis:

\begin{equation}
 {\rm cos}(\theta_0) = 1 + t_0/{(2k^2)}
 \label{semimajoraxisEq4}
\end{equation}

(The natural convergence domain of a series of Legendre polynomials is an ellipse with Foci at +1 and -1.)

 3) Polynomial bounded elastic amplitude and imaginary part thereof, $A(s,t)$:

\begin{equation}
 A(s,t) = 8 \pi \, s^{1/2} \, f(k,\theta) < {\rm when}\,( s \rightarrow
 \infty)  (s/(s_0))^{1+\Delta (t)}
\label{polyEq5}
\end{equation}

 We look for the maximal total cross section given by the optical theorem

\begin{equation}
   {\rm max}(\sigma_{\rm tot}) =  {\rm max}(2\pi/(k^2) \, \Sigma \, (2l+1) a_l)
\label{opttheoremEq6}
\end{equation}

Subject to (\ref{unitarityboundEq2}) and the inequality following
from (\ref{semimajoraxisEq4}) and (\ref{opttheoremEq6}):

\begin{equation}
A(s,t_0) = \Sigma \, a_l(s)(2l+1)P_l(1+t_0/(2k^2)) <
c(s/s_0)^{1+(\Delta (t_0)}
\label{AstoEq7}
\end{equation}

Both (\ref{AstoEq7}) and (\ref{opttheoremEq6}) are dominated by
high l waves. In the physical region, $-1 < cos(\theta) < 1$ , $  |P_l|<1$
and has l zeroes. The asymptotic behavior for large l is:

\begin{equation}
 P_l({\rm cos} (\theta))  \sim (\pi)l \, {\rm sin}(\theta/2))^{-1/2} \,
{\rm cos}
[(l+1/2)\theta]
\label{high1wavesEq8}
\end{equation}

For cos$(\theta) = 1+ 2t_0/s - $ or $ \, {\rm sin}(\theta) \sim \theta= i(4t_0/s)^{1/2}$
the  Legender polynomials grow exponentially with l:

\begin{equation}
   P_l(1+2t_0/s) \sim {\rm exp}((2l+1)(t_0/s)^{1/2})
\label{costhetaEq9}
\end{equation}

We omitted power-like pre-factors and used $2k^2 \sim s/2$ in the
large $s$ limit. Maximizing (\ref{opttheoremEq6}) subject to
(\ref{AstoEq7}) is equivalent to minimizing (\ref{AstoEq7}) when
(\ref{opttheoremEq6}) is held constant. The exponential growth of
the Pl's suggests using the lowest possible l waves. Consistent
with (\ref{unitarityboundEq2}) we then take

\begin{equation}
     a_l = 1 \; {\rm for} \, l < L ; \;\;a_l = 0 \; {\rm for} \; l > L.
\label{P1expgrowthEq10}
\end{equation}

The total cross section is then that of a black disc of radius
$R=(L/k)$ consisting of the geometric, inelastic cross section
$\pi R^2$ and matching shadow elastic:

\begin{equation}
   \sigma_{\rm tot} = 2\pi \, L^2/(k^2)
\label{shadowelasticmatchEq11}
\end{equation}

Substituting equation (\ref{P1expgrowthEq10}) for (the imaginary
parts of) the partial waves in (\ref{high1wavesEq8}):

\begin{equation}
 exp((2L+1)(t_0/s)^{1/2}) <  A(s,t_0) < c (s/(s_0)^{1+\Delta(t_0)}
\label{partialwavesEq12}
\end{equation}

(The left-hand side is a geometric series.  Approximating it
 by the last term introduces
only $ln(ln(s))$ corrections.)
Taking the log of (\ref{partialwavesEq12}) we find:

\begin{equation}
  L < k/((t_0)^{1/2}) \cdot [1+\Delta(t_0)] ln(s/(s_0))
\label{logof12EQ13}
\end{equation}

Substituting in Eq. (\ref{shadowelasticmatchEq11}) we finally
obtain the Froissart-Martin bound:

\begin{equation}
\sigma_{tot} (s) < 2\pi/(t_0) [1+\Delta(t_0)]^2 (ln(s/(s_0))^2
\label{FroissartMartinboundEq14}
\end{equation}

Historically this bound,  one of the very few rigorous result from  S matrix theory,  played an
 important role in excluding theories/models predicting cross-sections with power like rise
with energy.  The derivation utilized in addition to the usual of the  S matrix axioms also
 polynomial boundedness.  We are not aware however of any  model where this is not the case.
 Thus even Veneziano  amplitudes suggested as the ``Born Term'' in  ``String Like''-description
 of hadrons have linearly rising Regge trajectories  and for arbitrarily large time-like t
behave as $s^{\alpha(t)} $ with $\alpha(t)\sim \alpha(0) +t $ . However
for any finite $t_0$ the requirement of polynomial boundedness still holds.

 A power like rise in energy of $ A(s,t)$ and also of $\sigma(t)$ appears to arises in QCD  from the exchange
of gluon ladders in the t-channel  . However  unitarization/''eikonalization'' of such exchanges  yields a
 logarithmically expanding black disc and total  cross-sections  behaving like $ln^2 (s/s_0)$.
 If hadronic cross-sections indeed have a $ln^2(s)$ asymptotic behavior -as experimental data suggest-
 then the following question- the focus of the present work-as to what $t_0$ should appear in the F.B.- becomes meaningful.

 Formally the answer is clearcut: the lightest hadron exchangeable in the $t$ channel is the pion
 of mass $m$=140 MeV, and the nearest singularity in the imaginary part $A(s,t)$ is at $t_0= (2m)^2$.
 (By unitarity,  $A(s,t)$ involves products of two amplitudes and  requires  two-pion exchange).
 Yet we argue below that for energies W less than  14 TeV  (namely the LHC energy)
 a stronger "interim"  F.B. with the glue-ball mass being the nearest singularity, holds.

  pp collisions at the LHC are equivalent to  $10^{17}$ eV cosmic ray proton- fixed target collisions.
Cross-sections at higher energies are difficult to  obtain: The UHE cosmic ray spectrum
is cut- off at $E \sim 10^{20}$ eV and the flux the cosmic rays above $10^{17}$ eV is
extremely tiny making measurement of cross-sections very difficult.

For energies around LHC's $(s=(14,000)^2 \, GeV^2)$, existing cosmic ray data imply that
the F.B. with $s_0 \sim GeV^2$ exceeds the measured value of the inelastic $pp$ cross
sections of $\sim$ 80 mb by $\sim$ 100 so that it appears to be very far from being saturated.
However the following more detailed considerations are in order.

 The Froissart bound is universal applying to any pair $(a,b)$ of colliding hadrons. For $(a,b)$ different
 from p or$ {\bar p}$ we have data only for  $W< 40$ GeV. At such energies the cross-sections  depend on
 structural features which are specific to the colliding pair and vary from case to case: Two-versus three
 quarks in mesons and nucleons or the heavier $s$ quark in the Kaon (or $\phi$), yielding smaller $ {\bar q} q$
 bound states and smaller cross-sections of Kaons or of $\phi$ on prtons the cross-sections  of the  pions which are made of non-strange quarks.
Still all cross-sections clearly display a rise . In the PDG fit~\cite{PDG,Block}

\begin{equation}
\sigma^{a,b} = Z^{a,b}+Bln^2{(s/s_0)}+Y_1^{a,b}(1/s)^{\eta_1}+
Y_2^{a,b}(1/s)^{\eta_2} .
\label{PDGfitEq16}
\end{equation}

$\eta_{1,2}$ are the $t=0$ intercepts ($\sim$ 1/2) of the even and odd signature Regge trajectories-
-the contribution of the latter flipping sign between ${\bar a} b$ and $ab$ and the constant $Z^{ab}$
represents geometric, $a$- and $b$-dependent structural features.  For our purpose it is important that
in this best fit to existing data, the $Bln ^2 (s/s_0) $ term is universal  as expected  if the Froissart bound is  saturated.

Equating the fitted B=0.31 mb with $\pi/t_0$ leads to $t_0 \sim$
4.5 GeV$^2 \sim$ 55.4m$^2$. We suggest that the "effective" $t_0$
that can be used in the Froissart bound in the above energy regime
is $m_{\rm glue-ball}^2$--the (squared)  mass of the lightest
glue-ball. The glue-balls are heavier than the low-lying ${\bar q}q$
states. The lightest $0^{++}$ glue-ball mass computed~\cite{Lc} in lattice
QCD is ${\scriptstyle{> \atop\sim}} $1400 MeV and this is also the
mass of the lightest putative glue-ball candidate. Using it in $t_0$
reduces by a factor of 25  the margin with which the bound is satisfied
for the same $s_0$

 Several considerations motivate our suggestion.

i)${\bar q}q (q = u \, {\rm or} \, d)$ states such as $\pi (\rho)$,
etc., have  non-universal couplings--stronger to non-strange
hadrons and rather weak to hadrons, such as $ \phi(1020), \;
J/\psi$, etc., containing no u or d quarks and  hence their
exchange seems unlikely to control the universal Froissart bound.

ii) In 'tHooft's large $N_c$  limit or when all quark masses are
much larger than $\Lambda_{QCD}$, quark effects are turned off. The
universal F.B. applies also to glue-ball-glue-ball scattering and in the
large $N_c$ limit $t_0$ should pertain to the glue-ball sector only.

iii) High energy collisions are dominated by gluon physics as expected
from the enhanced Clebsch-Gordan coefficients in gluon $\rightarrow$ 2
gluons splitting function. This is indicated by Hera and Fermi Lab data
and a further dramatic enhancement expected at LHC makes it an intense
$gg$ machine.

iv) Spin-one gluon exchange dominates high energy collisions~\cite {Low,Nussinov1}.  It yields constant or,
 when iterated in the $t$ channel to form gluon ladders, cross sections rising as a  power of $s$ whereas
exchanging spin-half quarks or ${\bar q}q$ mesons/"Ordinary" Regge
trajectories yields the Y terms in Eq. (\ref{PDGfitEq16}), falling
like $\sim (1/s)^{1/2}$.

 iii) and iv) are valid in  pertrurbative QCD . There , and also with "BFKL" evolution--
 which builds the gluon exchange ladders in the $t$ channel--the gluons are massless
 and $1/(t_0) \sim(1/(m_g))^2$ yields a useless bound . Our suggested new bound is based on the
 assumption that this  high energy behavior of the  glue sector persists also non-perturbatively.
 We then expect the $t$ channel threshold- the lowest state in ``t channel'' cuts across the
 exchanged "Gluon blob", to have the mass of  the lightest glue-ball $m_{gb}$.

More specifically  the suggested stronger interim bound can be motivated  as follows:
The ordinary Froissart bound is derived for  a 2 $\rightarrow$ 2 amplitude $A(s,t)$ which has:

 a) a nearest (to 0) $t$ channel singularity at $t=t_0=4m^2 $  with m defined all along as the mass of the pion, {\it and} b)  grows  with energy like :  $A(s,t_0) \sim  s^{1+\Delta(t_0)}$.

 On first sight it seems that no such amplitudes exist: High energy amplitudes
 involving pion exchanges which naturally satisfy (a) with $t_0 = 4m^2$
 fail to satisfy (b).  Conversely, glue-ball exchange amplitudes guaranteed to satisfy
 (b) may not be singular at the two-pion threshold $t \sim t_0=4m^2$

 These issues are best illustrated in the context of Heisenberg's argument for the F.B.,
 predating Froissart by several years and which we briefly reproduce here in the framework of
 the renormalizable Yukawa model which existed already at that time:

\begin{equation}
 L_{Yukawa} = \int d^4 x \phi(x) \bar{\psi}_N(x)\psi_N(x).
\label{YukawaEq17}
\end{equation}

A bare nucleon at $r=0$ serves here as a point source for the pion's field: $\phi(r) =1/(r \, exp(mr))$.
When boosted to high energy along the z axis, say, it contracts but the ( Lortentz invariant)
profile still falls like $\exp-(mb)$ with impact parameter. The simplest version of the argument
utilizes only energy considerations and  ``geometry''  .

  The relevant interactions are between the target bare nucleon--at impact parameter B away and
 the contracted "disc-like" pion field of the projectile (and vice-versa) and some  interactions
 between the  pionic fields generated by the two nucleons. The maximal impact, $B_{max}$ for
 which non-vanishing inelastic collisions can occur when no transverse diffusion of the pionic field energy density
 is allowed,  is fixed as  follows:

We demand that the energy fraction of the projectile residing at impact parameters  $b>B_{max}$: $E\exp(-m_\pi \cdot B_{max})$ which  participates in the collision, exceed some threshold $E_{min}= s_0^{1/2}$
for producing the lightest hadronic system in the $s$ channel. This yields  the F.B. with $t_0$ and $s_0 \sim m^2$.
  It also implies a substantial part of  inelastic events  with just  one additional light hadronic state produced with big rapidity
 gaps to the two nucleons .

  Note that in order to saturate  the F.B we have to assume that elements from the two discs overlapping in
 impact parameter interact --no matter how large the rapidity gap between them .

 As we next show this assumption which undelies the above simple geometric argument fails for the Yukawa interactions of Eq. (\ref{YukawaEq17}):

To avoid self interactions we use the target's $\psi$ and the projectile's
$\phi$ and vice-versa. The boost contracts the $\phi$ disc in the
projectile, leaving the Lorentz invariant (pseudo) scalar
$\phi(x,t)$ the same. This yields a scattering amplitude $A(s,t)$
which is constant as a function of the energy $E$ failing to
satisfy (b). Indeed, this is what one expects from one $\pi$
exchange between the nucleons. It behaves for $s \rightarrow
\infty$ as $s^{J_{exchange}}$ and $J=0$ yields asymptotic constant
 imaginary part of forward amplitude and cross-sections falling like $1/s$

 However the geometric picture is justified in QCD !

A basic feature of QCD is that gluon and gluon ladder exchanges
automatically generate constant or rising cross sections. This is
true for one-gluon exchange between $q\bar{q}$ pairs and also if
the scattering hadrons were elongated chromoelectric flux tubes
confining the quarks at their ends. To see this let us repeat the above
discussion with $L(interaction) \sim E_1(x,t) \cdot E_2(x,t)$ the interaction effecting
 the scattering of particles 1 and 2. It follows from the Yang-Mills $F^2$ Lagrangian
if magnetic and self (11 and 22) interactions are emitted . We find that the simple
geometric picture where the overlapping portions of the flux tubes
have a finite probability to  to interact at all energies is now justified:~\cite{Nussinov2}.
 The vector nature of $A_\mu$ -or equivalently the boost invariant Gauss
law--implies that the z contraction of the chromoelectric flux tube by the
Lorentz  boost factor  $\gamma = (E/m)$ in the lab frame is compensated by
an increase by the same factor $\gamma$ of the chromoelectric fields  $E_x$ and
$E_y$--leading to $A(s,t)  \sim s$ and to constant cross sections.

The gluon exchange amplitudes involve the (induced) ${\bar q}-q$
color dipole moments or, for the confined bound states, the lengths of the
flux tubes and thus provide for a structure-dependent constant part of the
cross section naturally accounting for the constant $Z_{A,B}$ terms
 in the PDG fit of Eq. (\ref{PDGfitEq16}).

 Faster $s^{1+\Delta(0)}$ rising amplitudes can emerge if we iterate the
gluon exchanges in the $t$ channel--ala BFKL and /or other methods. The
saturation of gluon density and related issues are extensively discussed
in "low x" literature~\cite{Levin, Frankfurt, Mueller, Kovchegov}.  The key assumption that
we need to make in order to proceed further is  that the feature of gluon exchanges giving rise to $s^{1+D}$ rising
amplitudes $A(s,t)$ at time-like momentum transfers $t>0$  persists also non-perturbatively .With a glue like
blob exchanged in the t channel in high energy collisions the lowest t channel singularity,$ t_0$,
controlling the Froissart bound is $m_{gb}$. Also, we expect then that:  $ s_0 \sim m_{gb}^2$.

 Thus while gluon exchanges in QCD can account for the observed total and various inclusive cross-sections these
 exchanges do not naturally lead to $t_0= 4m^2$ ( with m the pion mass) in the F .B.
 To better understand this issue we recall some pre-QCD approaches to high energy scattering where  pions
 played a direct and dominant role.

Since as emphasized above, pion exchange cannot effectively couple hadrons having a large rapidity gap--
multi-peripheral models with repeated pion exchange in the $t$ channel were suggested some time ago~\cite{AFST}. In
parton model~\cite{Kogut} terms we have in the lab frame repeated evolution
of the projectile's partons or pions, with each splitting generating
a pion slower on average by some factor $ g$. After n steps with
$g^{-n}E <$ GeV, the last pion in the ladder can strongly interact
with the target. The ordering in rapidity of the successive ladder
rungs allowed summing the series yielding:

\begin{equation}
 A (s \rightarrow \infty, \;\; t \; {\rm fixed}) \sim \beta(t)\, s^{a(t)}.
\label{rapidityorderingEq18}
\end{equation}

  The universal $a(t)$ depends only on masses and on the couplings
of the $\pi\pi$ resonances produced in the multi-peripheral ladder.
 The approximation $a(t)= a(0) + a't$  implies a $a'ln(s)^{-1}$ shrinking
of the forward diffraction peak interpreted by Gribov as due to $\sim ln
(s)$ steps of a random walk in impact, b space, and the corresponding
increase by this factor of the (squared) range of interaction.

 Similar physics arises when the ($t$ channel) partial wave amplitude
$a(l,t)$ has a leading "Regge" pole~\cite{Regge} in the complex
angular momentum plane  (the one with largest real part) $\alpha(t)=a(t)$
and residue $\beta(t)$~\cite{Chew}.

  The  apparently constant asymptotic cross sections or $A(s,0) \sim s$ recquired
 within this framework  a "Pomeron" trajectory of intercept $\alpha_P(t=0)=1)$.
 Most particles lie on linearly rising trajectories of common slope  $d\alpha/dt \sim 1/{\rm GeV}^2$.

 However, the highest intercepts were $\sim$ 0.5 rather than the required 1, the slope
 of the Pomeron trajectory was much smaller than GeV$^{-2}$ and there were no spin-2
 particles lying on it. Since also the specific intercept of unity, while possible,
 is not guaranteed, the modern "Glue Pomeron" was adopted.

   $\pi\pi$ thresholds pervade much of hadronic physics and contribute to the the
 $g-2$ of the muon. Thus to find the mass of the $\rho$ meson one computes in lattice QCD the
 two-point $(x,y)$ function of vector currents and fits the $|(x-y)|$ dependence
with exp $ (-m_\rho|x-y|) $. For a stable $\rho$ this is indeed correct.
However, after a sufficiently long time  the two pions become the dominant
state and the asymptotic behavior reverts to exp$(-(2m|x-y|))$. By using $|x-y|$
 values larger than $m_\rho^{-1}$ yet smaller than $\tau$, the $\rho$ decay
 time, one can estimate $m_\rho$. The large ratio  $m_\rho /(\Gamma_ \rho)$
 required is provided  if $N_c>>1$. Likewise, the $\pi\pi$ admixture in
 $0^{++}$ glue-balls  only slightly modifies the masses of the glue-balls..

However, in the present case the space-like $|b|$, the analog of $|x-y|$, really tends
 to infinity, making pion thresholds far more relevant for the Froissart bound.

  A specific $A_c$ to be presented next has both $s^{(1+D)}$ rising amplitude and $exp(-2m.b)$
 behavior for large impacts. It does then provide a counter-example to the claim that no such
 amplitude exists and reinstates the original ``Axiomatic" F.B. with $ t_0=4m^2. $
 However the  small overall coefficient of $  A_c$  pushes to higher energies the onset of the
 original "axiomatic" F.B  and allows using our stronger, interim,  bound with $ t_0=m_{gb}^2$  for energies
 bellow W=14 TeV of LHC.

\begin{figure}
   \includegraphics[width=0.5\linewidth]{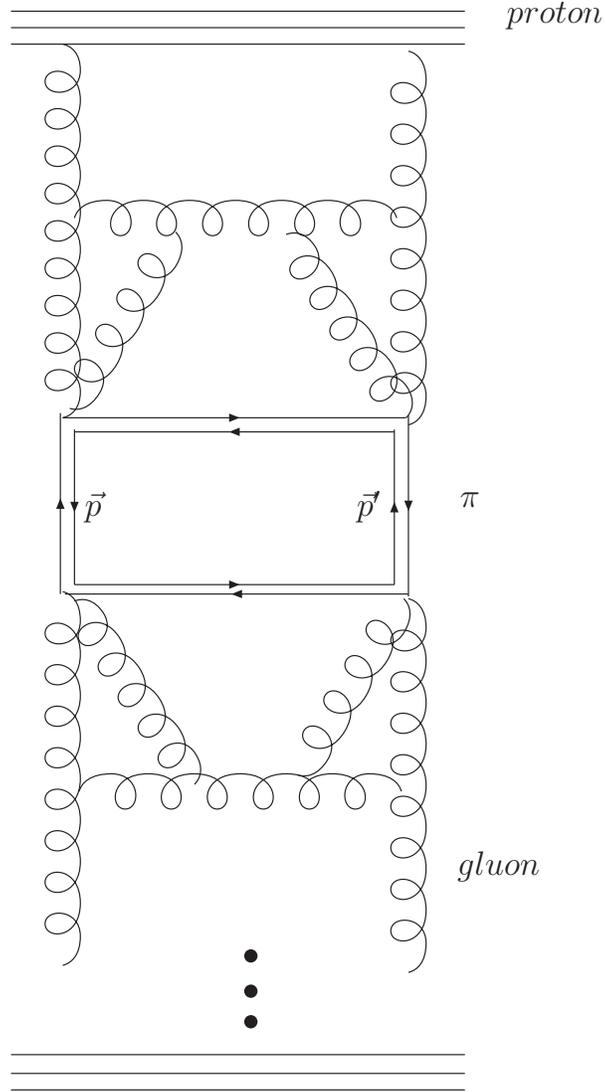}
  \caption{A generic diagram contributing to the amplitude $A_c$. The wavy lines describe
 gluons comprising  the upper and lower multi gluon ladders.  The  full double  lines  represent the pion circulating in the pion box .
 The latter couples to the gluonic blobs via color singlet gluon pairs at the four corners of the box .  }
\label{fig:gluon-pion}
\end{figure}

 The "hybrid" amplitude of \ref{fig:gluon-pion}  consists of gluon ladders with two-
 pion exchange insertion. The gluon ladders bridge most of the rapidity gap yielding the
 desired strong high energy behavior, $A \sim s^{(1+D)}$, and the pion pair with the slow
 b fall-off, $D_\pi(b) \sim  {\rm exp}(-(2mb))$, bridges most of the transverse coordinate space
 separation. Note that we need two pions to be exchanged in the $t$ channel whereas  two
(and multi) gluon states can couple to just  one $0^{++}$ glue-ball.

The Froissart bound with $t_0=4m^2$ clearly arises in a  $\lambda \pi^4$
theory with massive pions of  mass  m~\cite{Dashen, Veneziano}.

However,  the pions in QCD (and in the real world!) are pseudo-Goldstone Bossons associated with the
 spontaneous breaking of the global axial SU(2).

 The non-vanishing pion mass m  reflects the explicit symmetry breaking $u,d$ masses in
the Lagrangian: $m^2 \sim (m^0_u+m^0_d)$. The $(m^0_u \sim m^0_d {\rightarrow 0})$
symmetry limit should be smooth with no physical quantity changing dramatically.

 The reason is that confining QCD generates, independently of the quark
masses, a mass gap which serves as an infrared cutoff and avoids infrared
divergences of physical quantities. Hence cross sections cannot behave
like $m^{-2}$ and even asymptotic bounds with  overall coefficients of  $1/m^2$
seem puzzling.

 More generally any theory with no I.R. divergences the
  standard E.W. model or other field theories with no vanishing mass parameters,
 cannot develop such divergences even when spontaneous symmetry breaking yields
 massless Goldstone particles--a result known as Elitzurs' second theorem.

In view of this let us consider more closely the hybrid gluon ladder  with a nested  pion box to understand how-
all the abone arguements notwithstanding it does yield an  amplitude $A_c(s,t)$ which  gives rise to the standard
Froissart bound with $t_0=4m^2$.

The Goldstone character of the pions and the ensuing derivative
couplings in the low energy effective chiral Lagrangian imply that
the four coupling (to the upper and lower blobs in Fig. (gluon-pion) of
the exchanged pions vanish (have "Adler Zeroes") when the four momentums
 $p$ and $p'$ of the pions vanish.

The three momentums $p_i$ and $p'_i$ vanish at the $t$ channel threshold:
$t=t_0 = 4m^2$. More generally the space-like four momentums $p_\mu$ and $p'_\mu$ of the
two pions corresponding to transverse separations larger than $1/m$ between
the blobs in the figure, are smaller than $ m_{\pi}= m$ and vanish in
the m $\rightarrow 0 $ limit. Hence the couplings of the pions at these
four vertexes vanish  at this point  evading   $1/{(m^2)}$ divergences .
 This does not remove the square root branch point singularity at $t=t_0=4m^2$ for
 finite pion mass m but softens it.  The hybrid amplitude $A_c$  with the double-
 pion exchange is supressed at the two pion threshold in the t channel by $(m/M)^4$ due to
the four couplings at the corners where  M $\sim$ GeV is  the denominator mass in the momentum
-chiral perturbation- expansion. Further the introduction of (at least) two
quark loops yields $\sim  (2\pi \, N_c)^{-2}$ extra suppression. The last
suppression is present over and above the first chiral suppression, applying
if the two Nambu-Goldstone pions were replaced by any massive  meson made of a quark and an anti-quark.

 The amplitude with the two-pion exchange insertion is thus strongly suppressed
 relative to that of pure gluon exchange by a factor $F\sim 10^{-6}$ . More
 conservatively we estimate that F lies in the interval $ 10^{-8} -10^{-4}$ .
 Even with this large suppression factor the $A_c$ amplitude in s and b space,

\begin{equation}
 A_c(s,b) \sim F (s/s_0) \, {\rm exp}(-2mb)
\label{AcamplitudeEq19}
\end{equation}

 dominates for large b values the standard pure glue exchange amplitude:

\begin{equation}
 A_0(s,b) \sim  (s/s_0)) \, {\rm exp}(-m_{gb}b).
\label{puregluonamplEq20}
\end{equation}

  The apparent difficulty of saturating a F.B. with a $1/{m^2}$  coefficient  is resolved by the  $(m/M)^4$  factor included in the F  prefactor of  $A_c$ .
 This factor  pushes the crossover between the interim and assymptotic F.B to infinite s values as m approaches zero .

The full b profile for the collisions is the sum of the above two contributions: $ A(tot)=A_0+A_c $. The purely
 gluonic $ A_0$  tends to concentrate at small impact parameters ~$1/{m_{gb}}\sim 0.14$ Fermi and the hybrid -Glue + pion pairs-
diagram$ A_c$ has a five times  more extended profile $ 1/{2m}\sim 0.7$ Fermi but
a $ F \sim 10^{-6+/-2}$ times smaller coefficient.  The unitarity bound stating that the partial waves
(or the eikonal function)$ A (s,b)$ is less than unity,  should be applied to the full amplitude A(tot).
However the very different shapes of the two contribution allow the approximation
of applying this bound to each amplitude separately . Solving $A_0(s,b)\sim A_c(s,b)\sim 1$ then
yields an approximate value for the transition between the region of energies
where$ A_0 $dominates (and consequently our "interim" F.B. holds) and the region
where $A_c$ dominates and the "Asymptotic"  original F.B is reinstated :
 $ s/s_0 \sim F^{(1+2m/{m_{gb})}}~\sim 10^{7.2 +/- 2.4}$ and the
corresponding central b value is $ 2.2 +/- 0.8$ Fermi. We note that for $ s_0 \sim 20$
$GeV^2 $ symilar to the value used in the PDG fit,  the central value of the transition energy is
$3 \cdot 10^8 $$ GeV ^2$  just   above LHC's energy and the corresponding inelastic
black disc cross-section $\pi\cdot b^2\sim  150$  mb is less than a factor two higher than
that measured in cosmic ray experiments at these energies.

 This brings us to the  single most relevant question   : Is it concieveable that we
 will find at LHC energies signals of such a transition?

 The high rate of pp collisions allows, in principle,  measuring the total pp cross-sections
 at the nominal highest LHC energy of 14 TeV and perhaps also at  lower energies,  with unprecedented accuracy.
However, since even the more stringent interim F.B is not saturated at the LHC, this measurement  alone , may not indicate the above
 suggested transition.

 The blending in of the much larger and relatively transparent  "$A_c$ Disc" due to the pion box
 diagram could  generate a sharper diffraction peak at  small momentum transfers t.
 Unfortunately, the measurements  required to verify this are rather difficult at the LHC set-up.

 This still leaves us with the following  indirect possible indication. The calculations of cross-section for
 UHE CR proton - Air nucleus (and even more so for He- Air)  from pp cross-scetions~\cite{Block}~\cite{ Yodh}
  can be sensitive to the extended profiles~\cite{Maor} : the mutual shadowing is minimal for the extended, almost transparent,
  tails of the b profiles in  the$ A_c$ amplitude. Consequently the p-A/pp cross-section ratio may be considerabely larger
 than what computations based on Glaubers' theory with standard single Gaussian parameterization of the
 nucleons b space profiles would suggest  . The effect is even more  dramatic for He - Nucleus collisions since the compact Helium nucleus
 is compareable in size to the logarithmicaally expanding ``pion'' disc . It is tempting to speculate that the apparent indications in recent Auger data of
 component of the cosmic ray interactions occuring very high up in the atmosphere discrepancy is related to the enhanced proton -nucleus cross-sectionss.
  Conversely the  pp cross-
 section to be measured at LHC will then be lower than the values inferred  from cosmic ray data using the standard  analysis .\cite{Trefil}

  We close with the following remarks:

  I) The fact that to date, some 30-40 years after the introduction of QCD
 ~\cite{Greenberg, Nambu, Gell-Mann}  no glue ball states have been established
  impedes more quantitative implementation of our suggested 'interim" F.B.
  Finding the lightest glue-ball , presumably  a $ 0^{++}$ state as suggested by
 lattice calculations (but not proven to be so despite some theoretical efforts~\cite{West, Muzinich, Nussinov3})
  would fix the mass in the prefactor of the  interim Froissart bound.

  Also the partial decay width $\Gamma(gb\rightarrow\pi\pi)$ would help pin
 down the glue-ball  coupling to two pions in the amplitude $ A_c$ above allowing a
 more precise estimate of its suppression factor F,  and of the corresponding
 transition energy between the interim and asymptotic cross-sections.

 II.  several pieces of experimental data on pp and/or ${\bar p} p$ scattering at energies W up to
 2TeV indicate  that the Froissart bound is ${\it not}$ saturated in pp and/or ${\bar p} p$ scattering .
Thus $\sigma$(elastic)/$\sigma $(tot) is ~1/4 is significantly smaller than  1/2 as required for a
 black disc. Also the measured of slope of the differential cross-section
 at these energies  exceeds that predicted in a black disc model
 and  the differential cross-section does not have the $ J_1^2(Rt^{1/2})/t $ shape predicted for a black
 disc of radius R. Finally a black disc yields a purely imaginary elastic amplitude
 whereas the measured interference with the Coulomb real amplitude yields:
 $Re(A(s,t=0))/Im(A(s,t=0))\sim 0.1 $

 It is therefore gratifying that also our more strict F.B is not saturated.

  III.While the the growth of the proton proton crossection is  is  not far from saturating the interim
 F.B. we are  very far from the super asymptotic energies where
 {\it all} hadronic cross-sections starting with large nucleus nucleus
 cross-sections and ending with small $\Upsilon-\Upsilon$ cross-sections
 converge to the same asymptotic value~\cite{Frankfurt2}.
 Indeed as emphasized above the large yet relatively transparent disc due to the mixed -pion-gluon amplitude $A_c$
 tend to make proton deutirium crossection  grow initially  faster and only at super-assymptotic energies will
 both p-p and p-D crossections start converging to the comon large original large value of the Froissart limmit.
  We note that the universality of UHE hadronic scattering -expected when and if
  the F.B is fully and truly saturated - reflects not only in the independence
 of the value of the cross-sections on the incoming hadrons a and b. We also
 expect a universal (a,b) independent pattern of produced particles .
 These are just the various glue balls in our interim F.B. case and one
 additional low energy pi pi pair when the diagram $A_c$ 'kicks in" and we
 revert to the axiomatic F.B.
  This is suggested by s channel cuts across the corresponding diagrams.
 These cut (apart from the a and b fragmentation products at +/-Y(max))
  across just the gluonic blob or the latter and one  pair of pions
 in the two cases respectively. Unfortunately glue-balls are likely to
 be broad -decaying into many pions and  distinguishing the case when in addition to all
 the decay pions we  have another soft pair of pions at mid-rapidity  is  impossible.
 The universality of final states produced from the ( expanding) black disc is reminiscent
 of the universal spectrum of particles emitted via Hawking radiation from a black hole.
 In passing we recall that Heisenberg's argument for a (saturated)  Froissart bound  suggested
  a sizeable fraction of all final states consisting of a one pion pair with  large,  +/- Y(max) gaps to the elastic
 or diffracted  fragments of the incident protons . The "survival probability"
 of such large gaps was studied in Q.C.D. in connection with the production of just one
 Higgs particle at mid rapidity gap accompanied only by the  initial protons or
 some low lying diffractive excitation there-off  and is rather small ~\cite{Levin2}~\cite{Miller}
 The survival probability of the very light pion pair may be larger,but, just
 like for the forward diffraction peak will be extremely difficult to study experimentally.

  IV. To put Our interim ,yet stronger, F.B in prospective recall the  e.m ( photon exchange)
 contributions .To avoid Coulomb divergences  consider pn scattering.  At ultra high
 energies these are dominated by exchanges of ladders with  two photons coupling to closed
 electron boxes leading to the production  of many electron-positron pairs. The corresponding
 cross-sections have been calculated ~\cite{Wu} \cite{Brown} and as expected from the vanishing photon mass violate
  the ordinary Froissart bound ( with $t_0 \sim m_{\pi}^2$).
  The clear separation of strong and electro-magnetic interactions allows neglecting the latter at all
 foreseeable energies and the pre-factor $1/{t_0}$ in  the F.B.  remains $ (2m)^{-2}$.
  We suggest that the relation of our interim F.B with the large glue-ball
 denominator mass and the original F.B. with the pion mass is analogous  to
 that of the latter and the behavior expected when we have also the e.m.
 contributions . Here the analog of the massless photons are the gluons
 which due to the non-pertrurbative  QCD effects become massive
 and the the analog of the$ m_e\sim 1/2 MeV$ electron loops in the QED case are the light
 u/d ~ 4-8 MeV quarks manifesting here via the pion loop in $A_c$.  Finally
 the smallness of the EM couplings is reflected here in the rather large -chiral and
$ 1/N_c$ suppression of the pion loop in diagrams contributing to $ A_c$.
 While the disparity between this and the unsuppressed purely gluonic
 exchanges is much smaller than for EM versus hadronic processes - the
 notion that an interim more restrictive F.B can apply is similar .

  V. A recent treatment~\cite{Strassler} of the "QCD Pomeron" using an
 "ADS-CFT-Like" correspondence suggested a Froissart bound with
 a $1/{m_{gb}}^2$ pre-factor. Since the infinite $ N_c$ limmit is implicit in
 such approaches this is indeed expected.  Also the specific form of the F.B is natural from the
 ADS-CDT point-of view~\cite{Giddings}. The black disc is analog to
 a black hole -which sitting at a distance of ~$ln(s/{s_0})$ from the
 conformal AdS boundary can have a similar radius and a geometric
 cross-section ~$ ln(s/{s_0})^2$.  This is  further motivated by having  , as discussed
 in Sec III above ,emission of final state particles  in  ultra -high energy collisions,
 which is largely independent ( apart from  conserved Q.numbers) of the colliding particles.
 in analogy with to Hawking radiation  being independent of how the B.H is formed.

 I have been contemplating a stronger F.B. in the gluonic picture of the Pomeron for more than thirty years.
 Variants of this notion have already appeared sometime ago in , among others,  works by L.McLerran
 and by E.Gotsman  and (respective) collaboraors . The fact that such arguments fail in a pertrubative
 framework has been emphasized by A.Kovner.
   The final impetus for this work came during my recent visit to Ohio state University in a discussion with
 Anna Staso  .I am particularly grateful to Yuri Kovchegov for  patiently  listening to an earlier -rather primitive-version of this work ,
 for his most constructive criticism and for invaluable help in improving it . I also thank  C. Quigg and L. Frankfurt for drawing
my attention to the Auger puzzling results .

\end{document}